\documentclass[prd,showpacs,preprintnumbers,amsmath,amssymb,superscriptaddress,floatfix,nofootinbib]{revtex4}
\usepackage{graphicx}
\usepackage{epstopdf}
\usepackage{amsmath}
\usepackage{amsfonts}
\usepackage{amssymb}
\usepackage{bm}
\usepackage{braket}

\begin{document}
\title{Interpretation of the $\Omega_c \to \pi^+ \Omega(2012) \to \pi^+(\bar{K} \Xi)$ relative to $\Omega_c \to \pi^+\bar{K} \Xi$ \\ from the $\Omega(2012)$ molecular perspective}

\author{Natsumi Ikeno}
\email{ikeno@tottori-u.ac.jp}
\affiliation{Department of Agricultural, Life and Environmental Sciences, Tottori University, Tottori 680-8551, Japan}

\author{Wei-Hong Liang}
\email{liangwh@gxnu.edu.cn}
\affiliation{Department of Physics, Guangxi Normal University, Guilin 541004, China}
\affiliation{Guangxi Key Laboratory of Nuclear Physics and Technology, Guangxi Normal University, Guilin 541004, China}

\author{Genaro Toledo}
\email{toledo@fisica.unam.mx}
\affiliation{Instituto de Fisica, Universidad Nacional Autonoma de Mexico, AP20-364, Ciudad de Mexico 01000, Mexico}

\author{Eulogio Oset}
\email{oset@ific.uv.es}
\affiliation{Department of Physics, Guangxi Normal University, Guilin 541004, China}
\affiliation{Departamento de F\'{i}sica Te\'{o}rica and IFIC, Centro Mixto Universidad de Valencia - CSIC,
Institutos de Investigaci\'{o}n de Paterna, Aptdo. 22085, 46071 Valencia, Spain}

\date{\today}

\begin{abstract}
We present a mechanism for $\Omega_c \to \pi^+ \Omega(2012)$ production through an external emission Cabibbo favored weak decay mode, where the $\Omega(2012)$ is dynamically generated from the interaction of $\bar{K}\Xi^*(1530)$, $\eta\Omega$, with $\bar{K}\Xi$ as the main decay channel. The $\Omega(2012)$ decays latter to $\bar{K}\Xi$ in this picture, with results compatible with Belle data. The picture has as a consequence that one can evaluate the direct decay $\Omega_c^0 \to  \pi^+K^- \Xi^0$  and the decay $\Omega_c^0 \to  \pi^+\bar{K} \Xi^*$, $\pi^+\eta\Omega$ with direct coupling of $\bar{K}\Xi^*$ and $\eta\Omega$ to $K^- \Xi^0$. We show that, within uncertainties and using data from a recent Belle measurement, all these three channels account for about (12-20)\% of the total  $\Omega_c \to \pi^+K^- \Xi^0$ decay rate. The consistency of the molecular picture with all the data is established by showing that $\Omega_c \to \Xi^0 \bar{K}^{*0} \to \Xi^0K^- \pi^+$ together with $\Omega_c \to \pi^+ \Omega^* \to \pi^+K^- \Xi^0 $ account for about 85\% of the total $\Omega_c \to  \pi^+K^- \Xi^0 $.
\end{abstract}


\maketitle
\section{Introduction}
The discovery of the $\Omega(2012)$ by the Belle collaboration using $e^+e^-$ annihilation \cite{belle} prompted a fast and diverse reaction from the theoretical side, looking at it from the quark model perspective as the low-lying $p$-wave excited $J^P=3/2^-$ state \cite{zhong,aliev,azizi,polyakov,qifang, hosaka,qifangxie}, or as a molecular state stemming from the ${\bar K}\Xi^*(1530)$ and $\eta\Omega$ coupled channels interaction \cite{pavon,linzou,pavao,xiegeng,xiegeng2,toledo,liu}. There is much related work to $\Omega$ excited states in the literature which can be seen in the introduction of the references \cite{hosaka,qifangxie,toledo} and the recent review on the strange baryon spectrum in Ref.~\cite{hyodo}. Related work on new hadronic states can be seen in the recent review of Ref.~\cite{slzhu}.

We adopt the molecular perspective and in the present work we show the consistency of this picture with the latest result from Belle \cite{belle3}. The appealing feature of the molecular picture stems from the prediction of this state from the interaction of the ${\bar K}\Xi^*(1530)$ and $\eta\Omega$ channels in Refs.~\cite{hofmann,sarkar}. Even using quark models, in its version of the chiral quark model, a molecular structure of this type was claimed in the works of Refs.~\cite{zzye1,zzye2}. By using the Weinberg compositeness condition and a coherent sum of ${\bar K}\Xi^*$ and $\eta\Omega$, the molecular picture was also advocated in Ref.~\cite{thomas}.

The molecular picture was challenged by the Belle measurement of the $\Omega(2012) \to \pi{\bar K} \Xi$ and $\Omega(2012) \to {\bar K}\Xi$, showing a rate for the first decay channel of the order or less than 12\% of the second one \cite{belle2}. Since the 
$\pi{\bar K} \Xi$ channel is associated to ${\bar K} \Xi^*(1530)$, such a small fraction did not speak in favor of the ${\bar K} \Xi^*$ component. Yet, the phase space for this decay is very small and in Refs.~\cite{xiegeng2,toledo} the molecular picture was shown to be consistent with these data. 
A recent reanalysis of the Belle data, choosing also different cuts to determine the ${\bar K} \Xi^*$ content of the $\Omega(2012)$~\cite{newbelle} has established a value, rather than a boundary, for the $\pi \bar K \Xi$ to $\bar K \Xi$ ratio, which is 
\begin{equation}
 \mathcal{R}^{\Xi \pi \bar K}_{\Xi \bar K} = 0.97 \pm 0.24 \pm 0.07.
\label{eq:ratio_newbelle}
\end{equation}
The paper concludes that this ratio is ``consistent with the molecular interpretation for the $\Omega(2012)^-$ proposed in Refs.~\cite{pavon,pavao,xiegeng,thomas} (of Ref.~\cite{newbelle}), which predicts similar branching fractions for $\Omega(2012)^-$ decay to $\Xi(1530) \bar K$ and $\Xi \bar K$''. 
Indeed, in Table 5 of Ref.~\cite{pavao} the ratio of Eq.~(\ref{eq:ratio_newbelle}) ranges from 0.87 to 0.93 for different options, well within the range of the experiment.
In Refs.~\cite{xiegeng2,toledo}, the molecular picture was pushed to the limit and showed that the ratio $\mathcal{R}^{\Xi \pi \bar K}_{\Xi \bar K}$ could be made as small as 10\% but not smaller than that. We shall perform calculations with the parameters of Refs.~\cite{toledo} and~\cite{pavao} and compare the results with the experimental measurements of Ref.~\cite{belle3}.

More recently the $\Omega(2012)$ has been observed in the $\Omega_c$ decay \cite{belle3}. Anticipating the experiment on $\Omega_c$ decay, the authors of Ref. \cite{xiegeng3} studied the decay $\Omega_c \to \pi^+\Omega(2012) \to \pi^+(\pi {\bar K}\Xi)$ and $\Omega_c \to \pi^+\Omega(2012) \to \pi^+ {\bar K}\Xi$, and concluded that the decay mode $\Omega_c \to \pi^+\Omega(2012) \to \pi^+ {\bar K}\Xi$ was a good one to observe the $\Omega(2012)$ resonance, as was proved by its experimental measurement in Ref.~\cite{belle3}.
The new Belle measurements in Ref.~\cite{belle3} determine $\Omega_c \to \pi^+\Omega(2012) \to \pi^+ {\bar K}\Xi$ relative to $\Omega_c \to \pi^+ \Omega^-$ and $ \Omega_c \to \pi^+\Omega(2012) \to \pi^+ K^- \Xi^0$ versus $ \Omega_c \to \pi^+ K^- \Xi^0$ or $ \Omega_c \to \pi^+\Omega(2012) \to \pi^+ \bar{K}^0 \Xi^- $ versus $ \Omega_c \to \pi^+ {\bar K}^0 \Xi^-$. These ratios have been studied within a picture where the $\Omega(2012)$ is assumed to be the $1P$, $J^P=3/2^-$ excitation of the $\Omega$ in the quark model in Ref.~\cite{qifangxie} and consistency of the picture with data is found.
The conclusion is based on the consistency with the Belle measurements of Ref.~\cite{belle3}, Eqs.~(\ref{eq:ratio1}), (\ref{eq:ratio2}), but the ratio of Eq.~(\ref{eq:ratio_newbelle}) on which the Belle collaboration establishes the support for the molecular picture in Ref.~\cite{newbelle} is not discussed in Ref.~\cite{qifangxie}.

In the present work we wish to look at these data from the molecular perspective. What we do is to make a picture for the $ \Omega_c \to \pi^+\Omega(2012) \to \pi^+ {\bar K} \Xi$. Then we show that, associated to this signal within the molecular picture there is a background for $ \Omega_c  \to \pi^+ {\bar K} \Xi$, without going through the resonance, which does not belong to $ \Omega_c \to  {\bar K}^* \Xi \to \pi^+ {\bar K} \Xi $ reported in the PDG~\cite{pdg} nor to $\Omega_c \to \pi^+\Omega^* \to \pi^+ {\bar K} \Xi$, which is evaluated in Ref.~\cite{qifangxie} within the constituent quark model. The data put indeed a strong constraint on the background from these extra sources, which could be very large depending on the picture for the $\Omega(2012)$. What we find is that the new contribution for the $ \Omega_c \to   \pi^+ {\bar K} \Xi $ background is compatible with present data considering the known $ \Omega_c \to  {\bar K}^* \Xi$ and $\Omega_c  \to \pi^+\Omega^* $ sources.

\section{Formalism}\label{sec:form}
\begin{figure}[b]
 \includegraphics[width=0.33\linewidth]{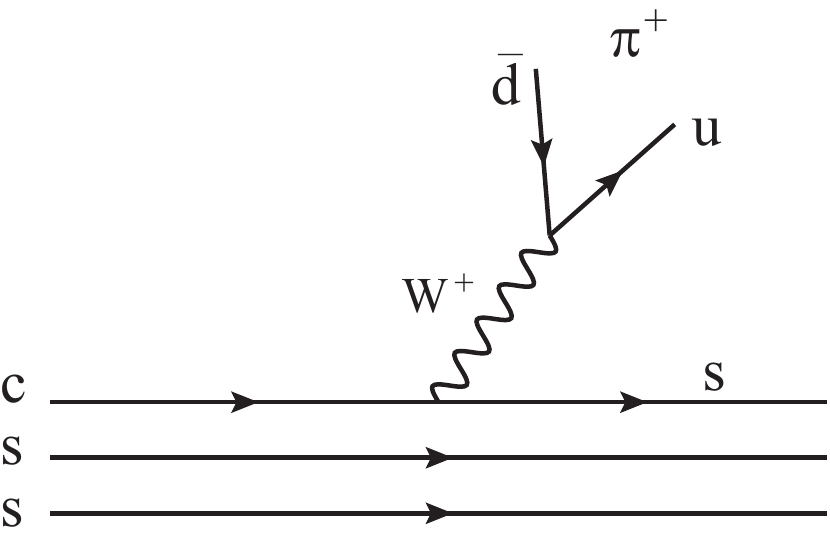}
 \caption{Microscopic picture for the weak decay $\Omega_c \to \pi^+ sss$ through external emission.}
\label{fig:1}
\end{figure}

In Ref.~\cite{belle3} the Belle collaboration reported on the ratios
\begin{equation}
  \frac{\mathcal{B}( \Omega_c^0 \to \pi^+ \,\Omega(2012)^-) \times  \mathcal{B}( \Omega(2012)^- \to  K^- \Xi^0)}{\mathcal{B}(\Omega_c^0 \to \pi^+ K^- \Xi^0)}
= (9.6 \pm 3.2 \pm 1.8) \%, 
\label{eq:ratio1}
\end{equation}

\begin{equation}
  \frac{\mathcal{B}( \Omega_c^0 \to \pi^+ \,\Omega(2012)^-) \times  \mathcal{B}( \Omega(2012)^- \to  \bar K^0 \Xi^-)}{\mathcal{B}(\Omega_c^0 \to \pi^+ \bar K^0 \Xi^-)}
= (5.5 \pm 2.8 \pm 0.7) \%.
\label{eq:ratio2}
\end{equation}
Let us see how this proceeds at the quark level. In Fig.~\ref{fig:1} we show the external emission mechanism for  $\Omega_c \to \pi^+ sss$.
The picture of Fig.~\ref{fig:1} offers a mechanism for $\pi^+ \Omega$, which is one important decay channel. Our decay channels require three particles in the final state, hence some hadronization must occur to produce the extra particle.

\begin{figure}[tb]
 \includegraphics[width=0.4\linewidth]{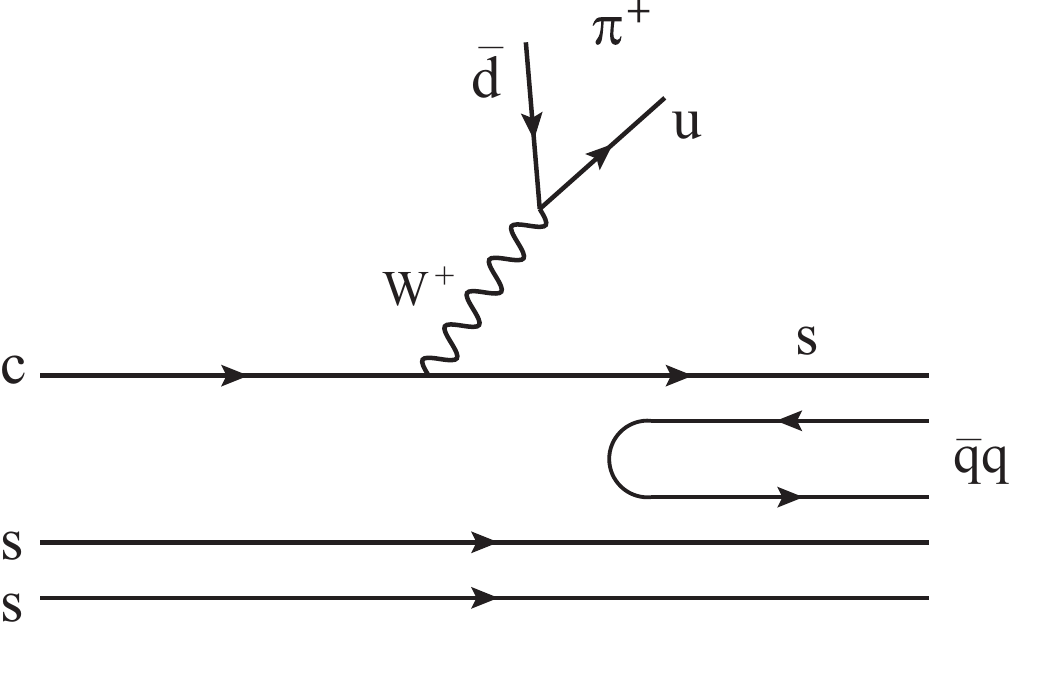}
 \caption{Hadronization of an $ss$ pair.}
 \label{fig:2}
\end{figure}

In the molecular picture for $\Omega(2012)$ one must create the building blocks $\bar K \Xi^*$ and $\eta \Omega$. Both of them have negative parity in $s$-wave which we consider. The one body mechanism of the picture of Fig.~\ref{fig:1} leaves the two lower $s$-quarks in the figure as spectators. We must hadronize the final state by introducing a $\bar q q$ pair with vacuum quantum numbers, and hence positive parity. Since parity is conserved after the weak vertex in Fig.~\ref{fig:1}, the upper $s$ quark in the final state must be produced in $L=1$ to have negative parity. Since finally we shall have the $s$ quark in the $\bar K$ in the ground state, the hadronization must involve this quark. In the $^3 P_0$ picture of hadronization~\cite{micu,yaouanc}, the $\bar q q$ pair is created with $L=1$ and $S=1$. Then the $L=1$ from the original $s$ quark combines with the second $L=1$ state to give zero orbital angular momentum, suited to the $\bar K$ and $\Xi$ (or $\eta, \Omega$) of the final state. The hadronization is depicted in Fig.~\ref{fig:2}.

After the hadronization the $sss$ state becomes
\begin{equation}
 sss \to \sum_{i} s\, \bar q_i q_i\, ss = \sum_i P_{3i}\, q_i\, s s, 
\end{equation}
where $P$ is the matrix of $q \bar q$ written in terms of pseudoscalar mesons 
\begin{equation}
 P \equiv 
\left(  
\begin{array}{ccc}
\frac{\pi^0}{\sqrt{2}}+\frac{\eta}{\sqrt{3}}+\frac{\eta^\prime}{\sqrt{6}}
 & \pi^+ & K^+ \\
 
\pi^- & -\frac{\pi^0}{\sqrt{2}}+\frac{\eta}{\sqrt{3}}+\frac{\eta^\prime}{\sqrt{6}} & K^0  \\
K^- & \bar{K}^0 & -\frac{\eta}{\sqrt{3}}+\sqrt{\frac{2}{3}}\eta^\prime
\end{array}    
\right) , 
\label{eq:Pmatrix}
\end{equation} 
where the $\eta-\eta'$ mixing of Ref.~\cite{bramon} is used. Thus, we obtain
\begin{equation}
 sss \to K^- uss + \bar K^0 d ss - \frac{\eta}{\sqrt{3}} sss,
\label{eq:sss}
\end{equation}
where we have neglected the $\eta'$ state which plays no role in the problem.

Next, we must find the overlap of the three quark states with the physical states $\Xi^0$, $\Xi^{-}$, $\Xi^{*0}$, $\Xi^{*-}$, $\Omega$. In terms of quarks these states are written as, 
\begin{equation}
 \Xi^0, \Xi^- : ~~ \phi = \frac{1}{\sqrt{2}} (\phi_{\rm MS} \, \chi_{\rm MS} + \phi_{\rm MA} \, \chi_{\rm MA}), 
\label{eq:WF_Xi}
\end{equation}
where $\phi_{\rm MS}$, $\phi_{\rm MA}$ are the mixed symmetric, mixed antisymmetric flavor wave functions and $\chi_{\rm MS}$, $\chi_{\rm MA}$ the mixed symmetric, mixed antisymmetric spin wave functions~\cite{close}. We have~\cite{miyahara,pavaojuan,debasliang}~\footnote{Note that consistency with the use of the phase convection implicit in Eq.~(\ref{eq:Pmatrix}) has as a consequence a different phase for the $\Xi^0$ state than in Ref.~\cite{close}. }
\begin{eqnarray}
\Xi^0:  \phi_{\rm MS} &=& \frac{1}{\sqrt{6}} [ s(us + su) - 2uss], \nonumber\\
        \phi_{\rm MA} &=& - \frac{1}{\sqrt{2}} [ s(us - su) ], 
\end{eqnarray}
\begin{eqnarray}
\Xi^-:  \phi_{\rm MS} &=& -\frac{1}{\sqrt{6}} [ s(ds + sd) - 2dss], \nonumber\\
        \phi_{\rm MA} &=&  \frac{1}{\sqrt{2}} [ s(ds - sd) ], 
\end{eqnarray}
\begin{eqnarray}
 \chi_{\rm MS} &=& \frac{1}{\sqrt{6}}(\uparrow\uparrow\downarrow + \uparrow\downarrow\uparrow - 2 \downarrow\uparrow\uparrow) ~~ {\rm for}~~ S_z=1/2  , \nonumber\\
 \chi_{\rm MA} &=& \frac{1}{\sqrt{2}}\uparrow (\uparrow\downarrow - \downarrow\uparrow)
 ~~ {\rm for}~~ S_z=1/2.
\end{eqnarray}
We shall only need the $S_z = 1/2$ as we shall see below. We shall start from $\Omega_c$ with $S_z = 3/2$, $\uparrow\uparrow\uparrow$, and then we only need the $\Xi^{*0}, \Xi^{*-}, \Omega$ with $S_z = 3/2$ and $S_z =1/2$, which we give below. For the $\Omega_c$ state we single out the heavy quark and symmetrize the light quark pair~\cite{isgur,roberts}.
Thus, we have
\begin{eqnarray}
& & \Omega_c  :~ css \uparrow \uparrow \uparrow \, , \nonumber\\
& & \Omega, S_z =3/2 :~ sss \uparrow \uparrow \uparrow \, ,\nonumber\\
& & \Omega, S_z =1/2 :~ sss \, \frac{1}{\sqrt{3}} (\downarrow \uparrow \uparrow + \uparrow \downarrow \uparrow + \uparrow \uparrow \downarrow   ) \, , \nonumber\\
& & \Xi^{*0}, S_z =3/2 :~ \frac{1}{\sqrt{3}}(sus + ssu +uss) \uparrow \uparrow \uparrow \, , \nonumber\\
& & \Xi^{*0}, S_z =1/2 :~ \frac{1}{\sqrt{3}}(sus + ssu +uss) \frac{1}{\sqrt{3}}  (\downarrow \uparrow \uparrow + \uparrow \downarrow \uparrow + \uparrow \uparrow \downarrow   ) \, ,  \nonumber\\
& & \Xi^{*-}, S_z =3/2 :~ \frac{1}{\sqrt{3}}(dss + sds +ssd) \uparrow \uparrow \uparrow \, , \nonumber\\
& & \Xi^{*-}, S_z =1/2 :~ \frac{1}{\sqrt{3}}(dss + sds +ssd) \frac{1}{\sqrt{3}}  (\downarrow \uparrow \uparrow + \uparrow \downarrow \uparrow + \uparrow \uparrow \downarrow   ) \, .
\label{eq:WF}
\end{eqnarray}

We take into account the weak interaction vertices. The $W \pi^+$ vertex is of the type~\cite{gasser,scherer}  
\begin{equation}
  \mathcal{L}_{W,\pi} \sim W^{\mu} \partial_{\mu} \phi.
\label{eq:L_wpi}
\end{equation}
Together with the $W q \bar q$ coupling of the type~\cite{chau,review}
\begin{equation}
  \mathcal{L}_{\bar qWq} \sim  (\bar q_{\rm fin} W_{\mu} \gamma^{\mu} (1-\gamma_5) q_{\rm in},
\label{eq:L_bwc}
\end{equation}
this leads to an interaction in the nonrelativistic approximation for the quark spins of the type~\cite{bayarliang}
\begin{equation}
 V_P = C(q^0 + \vec{\sigma} \cdot \vec{q}),
\label{eq:Vp}
\end{equation}
in the $\Omega_c$ rest frame~\cite{bayarliang}\footnote{We have redone the calculations in a different frame of reference, the one of $\bar K \Xi$ at rest, using the full relativistic expressions for the spinors discussed in Ref.~\cite{osetliang} and found only minor differences, smaller than the uncertainties from other sources, not worth being considered.}, where $q^0$ is the energy of the $\pi^+$ and $\vec q$ its three-momentum, and $C$ is a constant tied to the weak couplings and matrix elements of the radial wave functions of the baryon involved in terms of quarks.
We shall find a way to eliminate this constant in the analysis of our results by constructing appropriate ratios.

From Eq.~(\ref{eq:Vp}) we can see that with the $q^0$ operator we can make transition from $\Omega_c \uparrow \uparrow \uparrow$ to $\Xi^*$ or $\Omega$ in $S_z=3/2$ and with $\vec \sigma \cdot \vec q$ to $\Xi$ $(S_z = 1/2)$ or $\Xi^*$, $\Omega$ in ($S_z = 3/2, 1/2$). This assumes that the third component of the spin of the $s$ quark to the right of the $Wcs$ vertex in Fig.~\ref{fig:2} is the same as the one of the quark $q$ from the hadronization, which together with the $ss$ spectator quarks will make the $\Xi$, $\Xi^*$ or $\Omega$ final baryon. While this is intuitive since the $\bar K$ made from $s \bar q$ carries no spin nor angular momentum in the $s$-wave mode studied, we present a formal derivation in Appendix~\ref{App:A}.

\begin{figure}[tb]
 \includegraphics[width=0.4\linewidth]{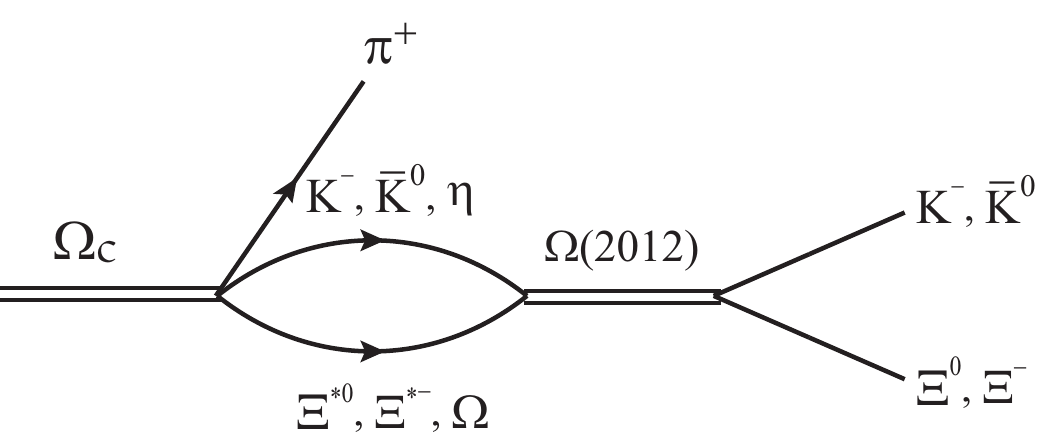}
 \caption{Mechanism for $\Omega_c \to \pi^+ \Omega(2012) \to \pi^+ \bar K \Xi$.}
 \label{fig:3}
\end{figure}

This said, the derivation of the matrix element of the $q^0 + \vec{\sigma}\cdot \vec{q}$ operator between the $\Omega_c$ original state and the final is straightforward. Defining 
\begin{equation}
 \vec{\sigma} \cdot \vec{q} = \sigma_+ q_- + \sigma_- q_+ + \sigma_z q_z,
\end{equation}
with
\begin{equation}
\sigma_+ = \frac{1}{2} (\sigma_x + i \sigma_y), ~~~
\sigma_- = \frac{1}{2} (\sigma_x - i \sigma_y),
\end{equation}
\begin{equation}
q_+ = q_x + iq_y ,~~~~~ q_- = q_x - iq_y,
\end{equation}
and taking into account the overlap of the states $uss$, $dss$, $sss$ of Eq.~(\ref{eq:sss}) with the flavor wave functions of the baryon states of Eqs.~(\ref{eq:WF_Xi}) and (\ref{eq:WF}) and the spin matrix operators of the $q^0 + \vec{\sigma} \cdot \vec{q}$ operator, we find the weight $W$ for the matrix elements of $\Omega_c \uparrow \uparrow \uparrow$ going to $\pi^+$ and the different final states as
\begin{eqnarray}
& & K^- \Xi^{*0} (S_z = 3/2):~~ W =  \frac{1}{\sqrt{3}} C (q^0 + q_z) \,, \nonumber\\
& & \bar K^0 \Xi^{*-} (S_z = 3/2):~~ W =  \frac{1}{\sqrt{3}} C (q^0 + q_z) \,, \nonumber\\
& & \eta \Omega (S_z = 3/2):~~ W = -\frac{1}{\sqrt{3}} C (q^0 + q_z) \,, \nonumber\\
& & K^- \Xi^{*0} (S_z = 1/2):~~ W = \frac{1}{3} C q_+ \, , \nonumber\\
& & \bar K^0 \Xi^{*-} (S_z = 1/2):~~ W = \frac{1}{3} C q_+  \,, \nonumber\\
& & \eta \Omega (S_z = 1/2):~~ W = -\frac{1}{3} C q_+  \,, \nonumber \\
& & K^- \Xi^{0} (S_z = 1/2):~~ W = \frac{\sqrt{2}}{3} C q_+ \, \nonumber\\
& & \bar K^0 \Xi^{-} (S_z = 1/2):~~ W = -\frac{\sqrt{2}}{3} C q_+  \, ,
\label{eq:Weight}
\end{eqnarray}
with $(q^0, \vec q\,)$ the four momentum of the $\pi^+$.
Note that in our notation we have the isospin doublets ($\bar K^0, -K^-$), ($\Xi^0, -\Xi^-$), and ($\Xi^{*0}$, $\Xi^{*-}$), so the $I=0$ states in the order of Ref.~\cite{sarkar} are
\begin{eqnarray}
| \Xi \bar{K}, I=0 \rangle &=& -\frac{1}{\sqrt{2}} \Big|\Xi^0 K^- - \Xi^- \bar{K}^0\Big\rangle\, ,\label{eq:XiK_I0}\\
| \Xi^* \bar K, I=0 \rangle &=& -\frac{1}{\sqrt{2}} \Big|\Xi^{*0} K^- + \Xi^{*-} \bar K^0 \Big\rangle\, \label{eq:XistarK_I0} .
\end{eqnarray}
There is a global arbitrary normalization, which is the same for all channels and which disappears in the ratios that we evaluate. The last two equations of Eq.~(\ref{eq:Weight}) allow the direct transition of $\Omega_c \to \pi^+ K^- \Xi^0$, $\pi^+ \bar K^0 \Xi^-$ without going through the $\Omega(2012)$ resonance. They contribute to the denominators in Eqs.~(\ref{eq:ratio1}) and (\ref{eq:ratio2}). On the other hand, the process $\Omega_c \to \pi^+ \Omega(2012) \to \pi^+ \bar K \Xi$ proceeds via the mechanism of Fig.~\ref{fig:3}, where one produces $\pi^+ \bar K \Xi^*$, $\pi^+ \eta \Omega$, the $\bar K \Xi^*$ and $\eta \Omega$ couple to the $\Omega(2012)$ resonance which later decay into $\bar K \Xi$.

\begin{figure}[tb]
 \includegraphics[width=0.45\linewidth]{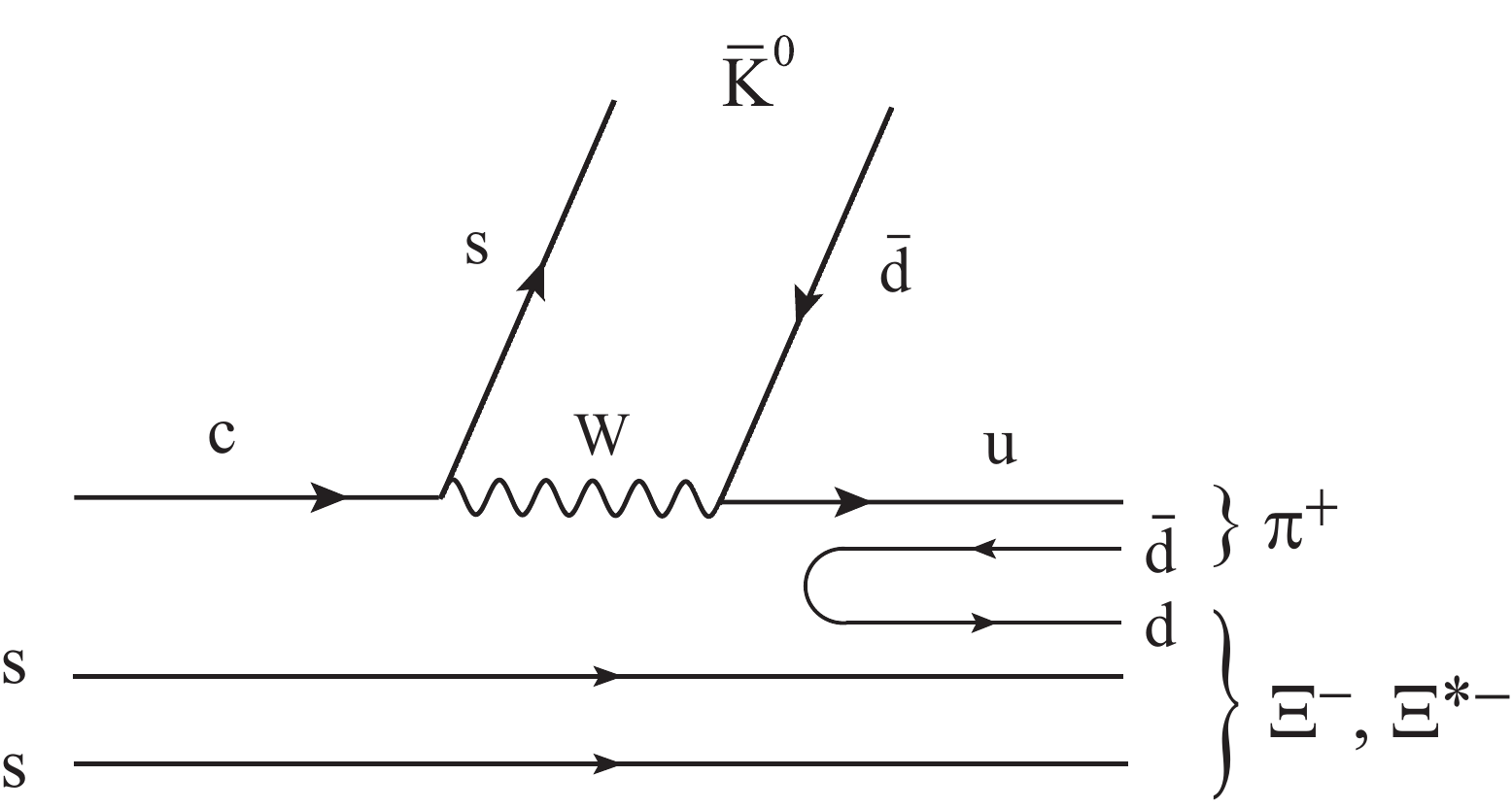}
 \caption{Mechanism for internal emission for $\bar K^0 \pi^+ \Xi^-$, $\bar K^0 \pi^+ \Xi^{*-}$.}
 \label{fig:4}
\end{figure}

The ratio of the width of the mechanism of Fig.~\ref{fig:3} to the background of $\Omega_c \to \pi^+ \bar K \Xi$ coming from all different sources should be compared to the ratios of Eqs.~(\ref{eq:ratio1}), (\ref{eq:ratio2}). We should note that in the case of Eq.~(\ref{eq:ratio2}) the final state is $\pi^+ \bar K^0 \Xi^-$. We can also have $\bar K^0 \pi^+ \Xi^-$, $\bar K^0 \pi^+ \Xi^{*-}$ production via internal emission as depicted in Fig.~\ref{fig:4}.
We can have $\bar K^0 \pi^+ \Xi^-$ production for the background through this mechanism although suppressed by a color factor around $\displaystyle \frac{1}{3}$. The $\bar K^0 \pi^+ \Xi^{*-}$ production can also contribute to the $\Omega(2012)$ production, but its projection into $I=0$ introduces an extra $\displaystyle \frac{1}{\sqrt{2}}$ factor in the amplitude, but the mechanism has a different structure as to external emission, and in the incoherent sum of external and internal emission we could expect corrections of the order of 6\%, We can accept larger uncertainties from this source, but in any case what is clear is that if we look at the $\pi^+ K^- \Xi^0$ final state production we just have external emission and the comparison of $\Omega(2012)$ signal and background production is much cleaner. The former argument also hints at a smaller ratio of signal to background for $\pi^+ \bar K^0 \Xi^-$ production than $\pi^+ K^- \Xi^0$ production, as seen in Eqs.~(\ref{eq:ratio1}), (\ref{eq:ratio2}), although the ratios could be compatible within errors. For all these reasons we shall only look at the ratio of Eq.~(\ref{eq:ratio1}).

\subsection{The $\Omega_c \to \pi^+ \Omega(2012) \to \pi^+ K^- \Xi^0$ reaction}
The process is depicted in Fig.~\ref{fig:3} and the amplitude for the process is given considering Eqs.~(\ref{eq:XiK_I0}), (\ref{eq:XistarK_I0}) by 
\begin{eqnarray}
 t &=& W(K^- \Xi^{*0}) \, G_{\bar K \Xi^*}(M_{\rm inv}) \, \left(-\frac{1}{\sqrt{2}}\right)\; g_{R, \bar K \Xi^*} \, \frac{1}{M_{\rm inv} - M_R + i \frac{\Gamma_R}{2} } \, \left( -\frac{1}{\sqrt{2}} \, g_{R, \bar K \Xi} \right) \nonumber \\
 &  & + W(\bar K^0 \Xi^{*-}) \, G_{\bar K \Xi^*}(M_{\rm inv}) \, \left(-\frac{1}{\sqrt{2}}\right)\; g_{R, \bar K \Xi^*} \, \frac{1}{M_{\rm inv} - M_R + i \frac{\Gamma_R}{2} } \, \left( -\frac{1}{\sqrt{2}} \, g_{R, \bar K \Xi} \right) \nonumber \\
 & & + W(\eta \Omega) \, G_{\eta \Omega}(M_{\rm inv}) \, g_{R, \eta \Omega} \, \frac{1}{M_{\rm inv} - M_R + i \frac{\Gamma_R}{2} } \, \left( -\frac{1}{\sqrt{2}} \, g_{R, \bar K \Xi} \right)
\nonumber \\
&=&  \left( -\sqrt{2} W(\bar K^- \Xi^{*0}) \, G_{\bar K \Xi^*}(M_{\rm inv}) \, g_{R, \bar K \Xi^*}  +  W(\eta \Omega) \, G_{\eta \Omega}(M_{\rm inv}) \, g_{R, \eta \Omega}
\right) 
\frac{1}{M_{\rm inv} - M_R + i \frac{\Gamma_R}{2} } \, \left( -\frac{1}{\sqrt{2}}\right) \, g_{R, \bar K \Xi} \,,
\end{eqnarray}
where the weights $W$ are given in Eq.~(\ref{eq:Weight}), $M_{\rm inv}$ stands for the invariant mass of the $K^- \Xi^0$ final state, and $R$, $M_R$, $\Gamma_R$, $g_{R, \bar K \Xi^*}$, $g_{R, \eta \Omega}$, $g_{R, \bar K \Xi}$ stand for the $\Omega(2012)$ resonance, its mass, width and couplings to $\bar K \Xi^*$, $\eta \Omega$ and $\bar K \Xi$. The functions $G_{\bar K \Xi^*}$ and $G_{\eta \Omega}$ are the loop functions of $\bar K \Xi^*$ or $\eta \Omega$ intermediate states, for which we use the cutoff regularized functions of Ref.~\cite{toledo}. The couplings of the resonance to the different channels are also taken from the work of Ref.~\cite{toledo}.

Next, we have to evaluate $|t|^2$ and sum over the polarizations of the $\Omega(2012)$. As we mentioned above, starting from $\Omega_c \uparrow  \uparrow  \uparrow$ we can reach $\Omega(2012)$ with $S_z =3/2$ and $1/2$. Then, considering the weights of Eq.~(\ref{eq:Weight}) we get
\begin{eqnarray}
\sum |t|^2 &=&  C^2 \left| \sqrt{2} G_{\bar K \Xi^*}(M_{\rm inv}) \, g_{R, \bar K \Xi^*}  +   G_{\eta \Omega}(M_{\rm inv}) \, g_{R, \eta \Omega}
\right|^2 
\nonumber\\& & 
\cdot \left( \frac{1}{3} \, q^{02} + \frac{1}{3} \, q^2_z + \frac{2}{3} \, q^0 q_z + \frac{1}{9} \, |q_+|^2
\right)
\frac{1}{2} \,
 \frac{| g_{R, \bar K \Xi} |^2}{ \left| M_{\rm inv} - M_R + i \frac{\Gamma_R}{2}\right|^2 } . 
\label{eq:t_res}
\end{eqnarray}
We can take the $z$ direction in the $\pi^+$ direction, and when integrating over the $\pi^+$ angles we shall get the angle averaged values of $q^2_z$, $q^0  q_z$ and $|q_+|^2$
\begin{equation}
\displaystyle  q^2_z \to \frac{1}{3} \, \vec{q}\,^2, ~~~~ q^0 q_z \to 0, ~~~~ |q_+|^2 = q^2_x + q^2_y \to \frac{2}{3} \, \vec{q}\,^2.
\nonumber
\end{equation}
Thus, $\displaystyle  \sum |t|^2 \to \sum |\bar t|^2$ with 
\begin{equation}
\frac{1}{3} \, q^{02} + \frac{1}{3} \, q^2_z + \frac{2}{3} \, q^0 q_z + \frac{1}{9} \, |q_+|^2
\to  \frac{1}{3} \, q^{02} + \frac{5}{27} \, \vec{q}\,^2.
\end{equation}
The $K^- \Xi^0$ mass distribution for the $\Omega_c$ decay is then given by 
\begin{equation}
 \frac{d \Gamma}{d M_{\rm inv} (K^- \Xi^0) } = \frac{1}{(2\pi)^3} \frac{M_\Xi}{M_{\Omega_c} } \, p_\pi \, \tilde{p}_{K^{-}} \sum |\bar t |^2 \,,  
\label{eq:dGam_Res}
\end{equation}
with 
\begin{equation}
 q = p_{\pi^+}= \frac{\lambda^{1/2}(M^2_{\Omega_c}, m^2_\pi, M^2_{\rm inv}(K^- \Xi^0) )}{2 M_{\Omega_c}},
\end{equation}
\begin{equation}
\tilde{p}_{K^{-}} = \frac{\lambda^{1/2}(M^2_{\rm inv}(K^- \Xi^0), m^2_{K^-}, M^2_{\Xi^0} )}{2 M_{\rm inv}(K^- \Xi^0) }.
\label{eq:pk}
\end{equation}

\subsection{Background for direct $\Omega_c \to \pi^+ K^- \Xi^0$}
In Eq.~(\ref{eq:Weight}) we also had the weights for $\Omega_c \to \pi^+ K^- \Xi^0$ direct transition.
Following the same argumentation as before we obtain also ${\rm d}\Gamma$ for this direct transition,
\begin{equation}
\label{eq:BG1}
 \frac{{\rm d} \Gamma^{(1)}_{\rm bac}}{{\rm d} M_{\rm inv}(K^- \Xi^0)}= \frac{1}{(2\pi)^3}\; \frac{M_{\Xi}}{M_{\Omega_c}} \; p_\pi\; \tilde{p}_{K^-}\; \sum\; \Big| \bar t\,^{(1)}_{\rm bac} \Big|^2,
\end{equation}
with $p_\pi, \tilde{p}_{K^-}$ as before and
\begin{equation}
\sum\; \Big|\bar t\,^{(1)}_{\rm bac} \Big|^2 = C^2 \frac{4}{27}\; \vec q^{\;2}.
\end{equation}

\subsection{Background for $\Omega_c \to \pi^+ K^- \Xi^0$ through intermediate $\bar K \Xi^*$ and $\eta \Omega$ states}
In Ref.~\cite{toledo} we had as coupled channels to obtain the $\Omega(2012)$,
$\bar K \Xi^*, \eta \Omega$ and $\bar K \Xi$,
the latter one in $D$-wave since $\Omega(2012)$ has $J^P=\frac{3}{2}^-$.
A fit to the data of Ref.~\cite{belle2} in Ref.~\cite{toledo} rendered
the coupling of $\bar K \Xi^*$ and $\eta \Omega$ to $\bar K \Xi$, all of them in $I=0$,
in terms of the parameters $\alpha$ and $\beta$:
\begin{equation}\label{eq:V}
  V_{\bar K \Xi^* \to \bar K \Xi} = \alpha\; \vec q_{\bar K}^{\;2}, ~~~~~~~
  V_{\eta \Omega \to \bar K \Xi} = \beta\; \vec q_{\bar K}^{\;2},
\end{equation}
with $\vec q_{\bar K}$ the $K^-$ momentum in the $K^- \Xi^0$ rest frame ($\tilde{p}_{\bar K}$ of Eq.~(\ref{eq:pk})).

\begin{figure}[b]
   \includegraphics[width=0.33\linewidth]{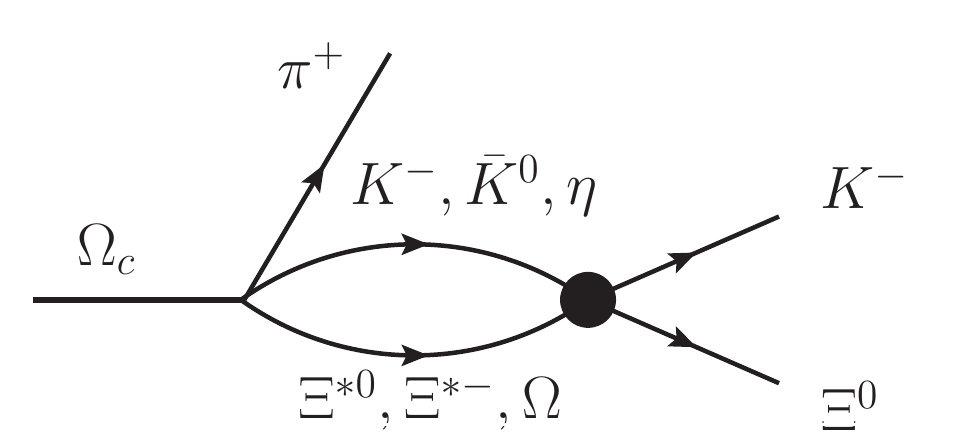}
    \caption{Mechanism for $\Omega_c \to \pi^+ K^- \Xi^0$ background production from intermediate $\bar K \Xi^*$ and $\eta \Omega$ states.}
    \label{fig:5}
\end{figure}

Then we have an additional mechanism to get background for $\Omega_c \to \pi^+ K^- \Xi^0$ as depicted in Fig.~\ref{fig:5}.
The amplitude for Fig.~\ref{fig:5} reads
\begin{eqnarray}\label{eq:tBG2}
 t^{(2)}_{\rm bac} &=& W(K^- \Xi^{*0})\cdot G_{\bar K \Xi^*}(M_{\rm inv})\cdot (-\frac{1}{\sqrt{2}})\; \alpha \;\vec p_{\bar K}^{\;2}\;(-\frac{1}{\sqrt{2}}) \nonumber \\[3mm]
 &&+ W(\bar K^0 \Xi^{*-})\cdot G_{\bar K \Xi^*}(M_{\rm inv})\cdot (-\frac{1}{\sqrt{2}})\; \alpha \;\vec p_{\bar K}^{\;2}\;(-\frac{1}{\sqrt{2}}) \nonumber \\[3mm]
&&+ W(\eta \Omega)\cdot G_{\eta \Omega}(M_{\rm inv})\cdot \beta \;\vec p_{\bar K}^{\;2}\;(-\frac{1}{\sqrt{2}}) \nonumber \\[3mm]
&=& W(K^- \Xi^{*0})\cdot G_{\bar K \Xi^*}(M_{\rm inv})\cdot \alpha \;\vec p_{\bar K}^{\;2}
-\frac{1}{\sqrt{2}}\;  W(\eta \Omega)\cdot G_{\eta \Omega}(M_{\rm inv})\cdot \beta \;\vec p_{\bar K}^{\;2}.
\end{eqnarray}
By using the same arguments as before and the values of $W$ of Eq.~(\ref{eq:Weight}),
we obtain a new source of background for $\Omega_c \to \pi^+ K^- \Xi^0$ given by
\begin{equation} 
\label{eq:BG2}
 \frac{{\rm d} \Gamma^{(2)}_{\rm bac}}{{\rm d} M_{\rm inv}(K^- \Xi^0)}= \frac{1}{(2\pi)^3}\; \frac{M_{\Xi}}{M_{\Omega_c}} \; p_{\pi^+}\; \tilde{p}_{K^-}\; \sum\; \Big|\bar t\,^{(2)}_{\rm bac} \Big|^2,
\end{equation}
with
\begin{equation}
 \sum \; \Big|\bar t\,^{(2)}_{\rm bac} \Big|^2 = \left| \alpha\; G_{\bar K \Xi^*}(M_{\rm inv})+\frac{1}{\sqrt{2}}\; \beta\; G_{\eta \Omega}(M_{\rm inv}) \right|^2 \cdot \tilde{p}^4_{K^-} \cdot 
C^2 \left( \frac{1}{3}\; {q^0}^2+ \frac{5}{27}\; \vec q^{\;2}  \right),
\end{equation}
where, as in Eq.~(\ref{eq:t_res}), we have summed over the $S_z = 3/2$ and $1/2$ components of $\bar K \Xi^*$ and $\eta \Omega$.

\section{Results}
In order to estimate uncertainties we use three sets of parameters $q_{\rm max}$,
that regularizes the $G_i$ loop functions,
$\alpha, \beta$, and the corresponding $g_{R, i}$ couplings and $\Gamma_R$ obtained in Ref.~\cite{toledo}.
The values of the parameters are tabulated in Table \ref{tab:1}~\footnote{We take advantage to mention that in the work of Refs.~\cite{hofmann,sarkar} the $\eta$ is considered as $\eta_8$, as in works of chiral perturbation theory. If one considers the $\eta$, $\eta'$ mixing of Ref.~\cite{bramon} the $C_{ij}$ coefficient of the potential between $\bar K \Xi^*$ and $\eta \Omega$, $C_{ij}=3$ becomes $3/\sqrt{2}$. We have redone the calculations with this option and the results change little since there is a trade off between the potential and the value of $q_{\rm max}$. The changes are smaller than the uncertainties from other sources discussed later on.}.
We also use the set of parameters of Ref.~\cite{pavao} which provides a ratio of  $\mathcal{R}^{\Xi \pi \bar K}_{\Xi \bar K}$ in very good agreement with the new Belle data~\cite{newbelle}.

\begin{table}[h]
\renewcommand\arraystretch{1.0}
\centering
\caption{Values of the parameters $q_{\rm max}, \alpha, \beta$ and the resulting $\Gamma_R$ ($R\equiv \Omega(2012)$) for three different sets of Ref.~\cite{toledo}~(sets~1--3) and the one of Ref.~\cite{pavao}~(set~4).} \label{tab:1}
\begin{tabular}{l|c|c|c|c}
\hline\hline
                            &    Set 1       &     Set 2    & Set 3     &  {Set 4}    \\
\hline
 $q_{\rm max} (\bar K \Xi^*)$~[MeV]~~     &  $735$         &  $775$        & $735$ & {$735$}  \\
$q_{\rm max} (\eta \Omega)$~[MeV]       &  $735$         & $710$         & $750$ & {$735$}  \\
 $\alpha$~[$10^{-8} \;{\rm MeV}^{-3}$]  & $-8.7$         &  $-8.7$       & $-11.0$  & {4.0} \\
 $\beta$~[$10^{-8}\; {\rm MeV}^{-3}$]   & $18.3$         & $18.3$        & $20.0$  & {1.5} \\
  $g_{R, \bar K \Xi^*}$                  &  $1.86-i0.02$  & $1.79+i0.02$  & $1.88+i0.04$  & {$2.01 +i0.02$ } \\
 $g_{R, \eta \Omega}$                   &  $3.52-i0.46$  & $3.79-i0.53$  & $3.55-i0.67$ & {$2.84 - i0.01$ }\\
 $g_{R, \bar K \Xi}$                    &~~~ $-0.42+i0.12$~~~  &~~~ $-0.44+i0.14$~~~ &~~~ $-0.42+i0.22$~~~ & {$~~~ -0.29+i0.04 ~~~$ } \\
 $\Gamma_{R}$ ~[MeV]                   & $7.3$  & $7.7$ & $8.2$ & {$6.24$} \\
\hline
\hline
\end{tabular}
\end{table}

Next, we define  ratios in which the unknown constant $C$ of the transition amplitude in Eq.~(\ref{eq:Vp}) cancels
\begin{eqnarray}
R_1 &\equiv& \dfrac{\Gamma^{(1)}_{\rm bac}}{\Gamma_{\rm signal}}
= \dfrac{\Gamma^{(1)}_{\rm bac}/ \Gamma_{\Omega_c}}{\Gamma_{\rm signal}/ \Gamma_{\Omega_c}}
=\dfrac{\mathcal{B}^{(1)}_{\rm bac}}{\mathcal{B}_{\rm signal}}, \label{eq:R1}\\[5mm]
R_2 &\equiv& \dfrac{\Gamma^{(2)}_{\rm bac}}{\Gamma_{\rm signal}}
= \dfrac{\Gamma^{(2)}_{\rm bac}/ \Gamma_{\Omega_c}}{\Gamma_{\rm signal}/ \Gamma_{\Omega_c}}
=\dfrac{\mathcal{B}^{(2)}_{\rm bac}}{\mathcal{B}_{\rm signal}}, \label{eq:R2}
\end{eqnarray}
where $\Gamma^{(1)}_{\rm bac}, \,\Gamma^{(2)}_{\rm bac}$ are obtained integrating the differential distributions of
Eqs.~\eqref{eq:BG1}, \eqref{eq:BG2} and $\Gamma_{\rm signal}$ is the integral of the mass distribution of Eq.~(\ref{eq:dGam_Res}) for the $\Omega_c \to \pi^+ \Omega(2012) \to \pi^+ K^- \Xi^0$ process.
Then we get the results for $R_1, R_2$ shown in Tables \ref{tab:2} and \ref{tab:3}.

\begin{table}[t]
\renewcommand\arraystretch{1.0}
\caption{Results for $R_1, R_2, R_1+R_2$ with different sets of parameters shown in Table \ref{tab:1}.} \label{tab:2}
\centering
\begin{tabular*}{0.45\textwidth}{@{\extracolsep{\fill}}cccc}
\hline\hline
         &  $R_1$     &  $R_2$    & $R_1 +R_2$   \\
\hline
 Set 1   &  $0.23$    &  $0.16$   & $0.39$  \\
\hline
 Set 2   &  $0.21$    &  $0.12$   & $0.33$  \\
\hline
Set 3    &  $0.21$    &  $0.20$   & $0.41$  \\
\hline
{Set 4}    &  {$0.45$}    &  {$0.17$}   & {$0.62$}  \\
\hline\hline
 \end{tabular*}
\renewcommand\arraystretch{1.0}
\centering
\caption{Results for $R_1, R_2, R_1+R_2$ with $\Gamma_R=6.4\; {\rm MeV}$ taken from PDG, and the other parameters shown in Table \ref{tab:1}.} \label{tab:3}
\begin{tabular*}{0.45\textwidth}{@{\extracolsep{\fill}}cccc}
\hline\hline
         &  $R_1$     &  $R_2$    & $R_1 +R_2$   \\
\hline
 Set 1   &  $0.20$    &  $0.14$   & $0.34$  \\
\hline
 Set 2   &  $0.17$    &  $0.10$   & $0.27$  \\
\hline
Set 3    &  $0.16$    &  $0.16$   & $0.32$  \\
\hline
{Set 4}    &  {$0.47$}    &  {$0.17$}   & {$0.64$}  \\
\hline\hline
\end{tabular*}
\end{table}

As we can see, the values of $R_1$ oscillate between $0.16-0.47$ while those of $R_2$ between $0.10-0.20$.
Altogether $R_1+R_2$ ranges between $0.27-0.64$.

From Eqs.~\eqref{eq:R1}, \eqref{eq:R2} we obtain
\begin{equation}
\label{eq:B1}
  \mathcal{B}^{(1)}_{\rm bac}=R_1 \; \mathcal{B}_{\rm signal},~~~~~~~~
  \mathcal{B}^{(2)}_{\rm bac}=R_2 \; \mathcal{B}_{\rm signal},
\end{equation}
and the global $\pi^+ K^- \Xi^0$ production branching fraction evaluated, equivalent to the magnitude of Eq.~(\ref{eq:ratio1}), but from the sources tied to the $\Omega(2012)$ production only, will be
\begin{equation}
\label{eq:Btot}
  \mathcal{B}[\Omega_c \to \pi^+ K^- \Xi^0]_{\rm th}= \mathcal{B}_{\rm signal}
                                             + \mathcal{B}^{(1)}_{\rm bac}
                                             + \mathcal{B}^{(2)}_{\rm bac}
  =\mathcal{B}_{\rm signal} (1+ R_1+R_2).
\end{equation}
Note that the ratios $R_1$, $R_2$ can be obtained in our approach as absolute numbers, since the unknown constant $C$ cancelled in the ratios, but the magnitude of Eq.~(\ref{eq:Btot}) implicitly includes the factor $C$ and we must find some experimental magnitude to eliminate it. For this purpose we proceed as follows.
If we divide Eq.~(\ref{eq:Btot}) by the experimental $\mathcal{B}[\Omega_c \to \pi^+ K^- \Xi^0]$,
we obtain
\begin{equation}
\label{eq:BRatio}
  \dfrac{\mathcal{B}[\Omega_c \to \pi^+ K^- \Xi^0]_{\rm th}}{\mathcal{B}[\Omega_c \to \pi^+ K^- \Xi^0]_{\rm exp}}
  =\dfrac{\mathcal{B}_{\rm signal}}{\mathcal{B}[\Omega_c \to \pi^+ K^- \Xi^0]_{\rm exp}}\;(1+ R_1+R_2).
\end{equation}
Our aim is to show that there is no inconsistency of the measured ratio of Eq.~(\ref{eq:ratio1})
and our hypothesis for the $\Omega(2012)$ formed from the interaction of
the $\bar K \Xi^*, \eta \Omega$ with the main decay channel as $\bar K \Xi$.
Then we take for $\dfrac{\mathcal{B}_{\rm signal}}{\mathcal{B}[\Omega_c \to \pi^+ K^- \Xi^0]_{\rm exp}}$ the ratio measured by Belle~\cite{belle2} (here is where the unknown constant $C$ is implicitly evaluated) and find
\begin{equation}
\label{eq:BRatio2}
  \dfrac{\mathcal{B}[\Omega_c \to \pi^+ K^- \Xi^0]_{\rm th}}{\mathcal{B}[\Omega_c \to \pi^+ K^- \Xi^0]_{\rm exp}}
  =(9.6\pm 3.2\pm 1.8)\;(1+ R_1+R_2) \%,
\end{equation}
which ranges from
\begin{equation}
\label{eq:BRatio3}
  (12.2-15.7)\%,
\end{equation}
with $38\%$ uncertainty from Eq.~\eqref{eq:BRatio2} summing errors in quadrature.
This means that we obtain with the three modes evaluated to produce $\pi^+ K^- \Xi^0$ only about $(12-16)\%$ of the total production of $\pi^+ K^- \Xi^0$,
with a maximum of about $20\%$ counting uncertainties.
This is a consequence of the molecular assumption made,
with the properties derived from it.
Note that both $R_1, R_2$ depend on $\mathcal{B}_{\rm signal}$ according to Eqs.~\eqref{eq:R1}, \eqref{eq:R2}
and we could have obtained in principle a fraction of $\pi^+ K^- \Xi^0$ from these sources
bigger than the experimental one,
showing a clear contradiction.
In what follows we show that, indeed,
we should not get a bigger fraction of $\pi^+ K^- \Xi^0$ production from our sources
because there are two other sources not counted by us,
which account for more than $80\%$ of the total $\pi^+ K^- \Xi^0$ production.

One of the sources not included by us in $\Omega_c \to \pi^+ K^- \Xi^0$
comes from $\Omega_c \to \Xi^0 \bar K^{*0} \to \Xi^0 K^- \pi^+$ ($\Gamma_8$  of PDG~\cite{pdg}).
We can see that
\begin{equation}
\label{eq:pdg}
  \dfrac{\Gamma_8}{\Gamma_7}= \dfrac{\mathcal{B}[\Xi^0 \bar K^{*0} \to \Xi^0 K^- \pi^+]}{\mathcal{B}[\Xi^0 K^- \pi^+]}
  =\dfrac{0.68\pm 0.16}{1.20\pm 0.18} =0.57\pm 0.16,
\end{equation}
where we have added relative errors in quadrature.
This means that about $60\%$ of the $\Omega_c \to \Xi^0 K^- \pi^+$ decay comes from the $\Xi^0 \bar K^{*0} \to \Xi^0 K^- \pi^+$ decay
which is not a part of our calculation (we cannot relate it to the sources studied in our formalism).
There is another source of $\Xi^0 K^- \pi^+$ production.
With the mechanism of Fig.~1 one can produce $\pi^+ \Omega^*$
where $\Omega^*$ are any kind of excited $\Omega(sss)$ states.
The posterior decay of $\Omega^* \to \Xi^0 K^-$ would be another source of $\Xi^0 K^- \pi^+$ production
not tied to the mechanism evaluated by us.
There is no information on such decay in the scarce information on $\Omega^*$ states in the PDG,
but we can rely upon a theoretical quark model calculation for an approximate estimate of such contributions.
The answer to this question is provided in Ref.~\cite{qifangxie}.
There we find
\begin{equation}
\label{eq:BrXie1}
  \dfrac{\mathcal{B}[\Omega_c \to \pi^+ \,\Omega(1\,^2P_{3/2^-}) \to \pi^+ K^- \Xi^0]}{\mathcal{B}[\Omega_c \to \pi^+ K^- \Xi^0]}\simeq 0.08,
\end{equation}
\begin{equation}
\label{eq:BrXie2}
  \dfrac{\mathcal{B}[\Omega_c \to \pi^+ \, \Omega(1\,^2P_{1/2^-}) \to \pi^+ K^- \Xi^0]}{\mathcal{B}[\Omega_c \to \pi^+ K^- \Xi^0]}\simeq 0.11,
\end{equation}
\begin{equation}
\label{eq:BrXie3}
  \dfrac{\mathcal{B}[\Omega_c \to \pi^+\, \Omega(1\,^4D_{1/2^+}) \to \pi^+ K^- \Xi^0]}{\mathcal{B}[\Omega_c \to \pi^+ K^- \Xi^0]}\simeq 0.04,
\end{equation}
\begin{equation}
\label{eq:BrXie4}
  \dfrac{\mathcal{B}[\Omega_c \to \pi^+ \,\Omega(1\,^4D_{3/2^+}) \to \pi^+ K^- \Xi^0]}{\mathcal{B}[\Omega_c \to \pi^+ K^- \Xi^0]}\simeq 0.02.
\end{equation}
Summing all these contributions we find a fraction of
\begin{equation}
\label{eq:BrXieSum}
  \dfrac{\mathcal{B}[\Omega_c \to \pi^+ \,\Omega^* \to \pi^+ K^- \Xi^0]}{\mathcal{B}[\Omega_c \to \pi^+ K^- \Xi^0]}\simeq 0.25.
\end{equation}
If we sum this fraction to the one of Eq.~\eqref{eq:pdg} we already find $100\%$ of the $\Omega_c \to \pi^+ K^- \Xi^0$ decay.
Counting errors,
without considering the unknown ones from
Eqs.~\eqref{eq:BrXie1}, \eqref{eq:BrXie2}, \eqref{eq:BrXie3}, \eqref{eq:BrXie4}, we find a total fraction of
\begin{equation}
\label{eq:tot}
  0.82\pm 0.16 \simeq 66-98\%.
\end{equation}
The margin should be bigger considering uncertainties from the quark model,
and this result matches well with the $(12-16)\%$ with $38\%$ uncertainty fraction
obtained in Eq.~\eqref{eq:BRatio3} from sources related to the $\Omega(2012)$ considered as a molecular state.
We should stress that, should we have obtained a much smaller $\mathcal{B}_{\rm signal}$ than the one we obtained,
the ratios obtained for $R_1$ and $R_2$ would have been much bigger,
such as to make inconsistent the molecular picture of the $\Omega(2012)$.
We certainly cannot see the results found as a proof of the molecular nature of the $\Omega(2012)$, but we see that the picture is consistent with this valuable experimental data through the nontrivial test done.

\section{Conclusions}
We have carried out calculations of the $\Omega_c \to \pi^+ \Omega(2012) \to \pi^+ K^- \Xi^0$ decay
from the perspective that the $\Omega(2012)$ is a molecular state build up from the $\bar K \Xi^*(1530), \eta \Omega$ channels which decay mostly in the $\bar K \Xi$ channel.
The process proceeds via external emission with a Cabibbo favoured mode,
resulting in $\pi^+$ emission and the formation of a baryon system with $sss$ quarks.
We allow for the hadronization of an $ss$ pair, leading to $\bar K \Xi^*, \eta \Omega$, which interact giving rise to the $\Omega(2012)$ which later decays to $K^- \Xi^0$. However, the same operator from the weak transition and hadronization can lead to direct $K^- \Xi^0$ production,
hence contributing to a background process in $\Omega_c \to \pi^+ K^- \Xi^0$ decay.
This means that we can relate these processes.
At the same time, the analysis of the $\Omega(2012)$ in Ref.~\cite{toledo}
with three channels $\bar K \Xi^*, \eta \Omega$ and $\bar K \Xi$ leads to the coupling of $\bar K \Xi^*, \eta \Omega$ to $\bar K \Xi$
such that when we produce the $\bar K \Xi^*, \eta \Omega$ in the first step,
we can go from these states to $\bar K \Xi$ without passing through the resonance,
providing another source of background for $\Omega_c \to \pi^+ K^- \Xi^0$.
The three processes are tied to the properties of the $\Omega(2012)$ as a molecular state,
consistent with the Belle data~\cite{belle2} and their ratios are very sensitive to these properties.
We find that counting uncertainties, the three sources of $\Omega_c \to \pi^+ K^- \Xi^0$ decay evaluated account for about $(12-20)\%$ of the total $\Omega_c \to \pi^+ K^- \Xi^0$ branching ratio,
but this quantity could be significantly larger should we have a different molecular picture with different couplings to the building channels.
The consistency with experiment of this obtained fraction is established
when we show that the $\Omega_c \to \Xi^0 \bar K^{*0} \to \Xi^0 K^- \pi^+$ together with $\Omega_c \to \pi^+ \Omega^* \to \pi^+ K^- \Xi^0$ decay channels account for about $85\%$ of the total $\Omega_c \to \pi^+ K^- \Xi^0$ decay, as obtained in a quark model.
We could show that the ratio of Eq.~(\ref{eq:ratio1}) from the Belle experiment~\cite{belle3} has been very useful to establish a new test for the nature of the $\Omega(2012)$ and look forward to other measurements that can put further constraints to the different pictures of the $\Omega(2012)$.

\appendix
\section{Microscopical evaluation of the hadronization using the $^3P_0$ model}\label{App:A}
 We follow the steps of section 3 of Ref.~\cite{bayarliang}. We assume the interaction of zero range as done in Ref.~\cite{brown} to explain the phenomenon of pairing in nuclei, and we  worry only about angular momentum and spin matrix elements. The $s$ quark to the right of the weak vertex in Fig.~\ref{fig:2} has $L=1$ and thus couples to $\ket{J,M}$,
\begin{equation}
\ket{J,M}=\sum_m {\mathcal C}(1 \ 1/2 \ J; m, \ M-m) Y_{1m} \ket{1/2, \ M-m}.
\label{A1}
\end{equation}
On the other hand, the ${\bar q}q$ from hadronization has $L^\prime=1$ and $S=1$ coupling to total angular momentum zero. The spin state is $\ket{1S_3}$ given by (we follow nomenclature and formulas of Ref.~\cite{rose})
\begin{equation}
\ket{1S_3 }=\sum_s {\mathcal C}(1/2, \ 1/2 \ 1; s, \ S_3-s) \ket{1/2, \ s} \ket{1/2, \ S_3-s},
\label{A2}
\end{equation}
and the global angular momentum state is
\begin{eqnarray}
\ket{0 0}&=&\sum_{M_3,S3} {\mathcal C}(1 \ 1 \ 0; M_3, S_3, 0) Y_{1M_3}\ket{1S_3}\nonumber \\
&=&\sum_{M_3} {\mathcal C}(1 \ 1 \ 0; M_3, -M_3) Y_{1M_3} \cdot \nonumber\\
&\cdot& \sum_{s} {\mathcal C}(1/2 \ 1/2 \ 1; s, -M_3-s )\ket{1/2, s} \ket{1/2, \ -M_3-s}.
\label{A3}
\end{eqnarray}

The $Y_{1m}$ of Eq. (\ref{A1}) and  $Y_{1M_3}$ of Eq. (\ref{A3}) combine to $L^{\prime \prime}=0$ and we have \cite{rose}
\begin{equation}
 Y_{1m} Y_{1M_3}  \to (-1)^m \frac{1}{4\pi} \delta_{M_3,-m}\nonumber .
\end{equation}
 Then, using ${\mathcal C}(1 \ 1 \ 0; M_3, -M_3, 0)=(-1)^{1-M_3} \frac{1}{\sqrt{3}}= (-1)^{1+m} \frac{1}{\sqrt{3}}$
 \begin{eqnarray}
\ket{J M} \ket{0 0}&=&- \frac{1}{\sqrt{3}} \frac{1}{4\pi} \sum_{m}\sum_{s} {\mathcal C}(1 \ 1/2 \ J; m, M-m) \nonumber \\
&\cdot& {\mathcal C}(1/2 \ 1/2 \ 1; s, m-s) \ket{1/2, M-m}\ket{1/2, s} \ket{1/2, \ m-s} \nonumber.
\end{eqnarray}
But now $\ket{1/2, M-m} \ket{1/2, \ s} $ combine to give the ${\bar K}$, hence they couple to $j=0$
 \begin{eqnarray}
\ket{1/2, M-m} \ket{1/2, s}& \to&  {\mathcal C}(1/2 \ 1/2 \ 0; M-m, s)\ket{0,0 ({\bar K}) }\nonumber \\
&=& (-1)^{1/2+s} (1/2)^{1/2} \ket{0,0 ({\bar K})}\nonumber,
\end{eqnarray}
and thus
\begin{eqnarray}
\ket{J M} \ket{0 0}&=&- \frac{1}{\sqrt{3}} \frac{1}{4\pi} \frac{1}{\sqrt{2}} \sum_{s} (-1)^{1/2+s} {\mathcal C}(1 \ 1/2 \ J; M+s, -s) \nonumber \\
&\cdot& {\mathcal C}(1/2 \ 1/2 \ 1; s, M) \ket{0,0 ({\bar K}) }  \ket{1/2, \ M} \nonumber .
\end{eqnarray}
Permuting the order of the arguments in the Clebsch-Gordan coefficients \cite {rose} we get
\begin{eqnarray}
\ket{J M} \ket{0 0}&=&- \frac{1}{\sqrt{6}} \frac{1}{4\pi} \sum_{s} (-1)^{1/2+s}(-1)^{1/2-s} (3/2)^{1/2} {\mathcal C}(1 \ 1/2 \ J; M+s, -s) \nonumber \\
&\cdot& {\mathcal C}(1 \ 1/2 \ 1/2 ; M+s, -s) \ket{0,0 ({\bar K}) }  \ket{1/2, \ M} \nonumber\\
&=& \frac{1}{2} \frac{1}{4\pi}\ket{0,0 ({\bar K}) }  \ket{1/2, \ M} \delta_{J,1/2} \nonumber \, .
\end{eqnarray}
This means that, after removing the $\bar{K}$ from the $s\bar{q}q$ system, what remains is a spin 1/2 state with third component $M$, the third component of the total spin of the $s$ quark to the right of the weak vertex. This is the state with which we have to make transitions with the $q^0 + \vec{\sigma} \cdot \vec{q} $ operator between the original $c$ quark and the $q$ quark of the final baryon.

\section*{ACKNOWLEDGMENT}
The work of N. I. was partly supported by JSPS KAKENHI Grant Number JP19K14709. 
This work is  partly supported by the National Natural Science Foundation of China under Grants No. 11975083 and No. 12147211.
G.~T. acknowledges the support of DGAPA-PAPIIT UNAM, grant no. IN110622.
This work is also partly supported by the Spanish Ministerio de Economia y Competitividad
and European FEDER funds under Contracts No. PID2020-112777GB-I00
and by Generalitat Valenciana under contract PROMETEO/2020/023.
This project has received funding from the European Unions Horizon 2020 research and innovation programme
under grant agreement No. 824093 for the ``STRONG-2020'' project.

\end{document}